\documentclass[a4paper,11pt]{article}
\pdfoutput=1 

\usepackage{jheppub} 

\usepackage[T1]{fontenc} 

\def\beq{\begin{eqnarray}}
\def\eeq{\end{eqnarray}}
\newcommand{\nn}{\nonumber}

\def\Det{\,\mbox{Det}\,}

\DeclareMathOperator{\cx}{\square}

\def\al{\alpha}
\def\be{\beta}

\def\ga{\gamma}
\def\de{\delta}

\def\ep{\epsilon}

\def\ka{\kappa}
\def\la{\lambda}

\def\pa{\partial}

\def\ph{\varphi}

\def\th{\theta}

\def\La{\Lambda}

\def\Th{\Theta}

\usepackage{soul} 

\def\cred{\color{black}}

\title{\boldmath Bound states of massive complex ghosts in superrenormalizable quantum gravity theories}

\author[a,1]{M.  Asorey,\note{Corresponding author.}}
\author[b]{G. Krein,}
\author[a]{M. Pardina,}
\author[c]{I. Shapiro}

\affiliation[a]{Centro de Astropart\'{\i}culas y F\'{\i}sica de Altas Energ\'{\i}as,
Departamento de F\'{\i}sica Te\'orica.
\\
Universidad de Zaragoza, E-50009 Zaragoza, Spain}
\affiliation[b]{Instituto de F\'{\i}sica Te\'orica, Universidade Estadual
Paulista, Rua Dr. Bento Teobaldo Ferraz, 271 - Bloco II, 01140-070
 S\~ao Paulo, SP, Brazil}
\affiliation[c]{Departamento de F\'{\i}sica, ICE,
Universidade Federal de Juiz de Fora,
Campus Universit\'{a}rio,
\\
 Juiz de Fora, 36036-900, MG, Brazil}

\emailAdd{asorey@unizar.es}
\emailAdd{gastao.krein@unesp.br}
\emailAdd{mpardina@unizar.es}
\emailAdd{ilyashapiro2003@ufjf.br}

\abstract{One of the remarkable differences between renormalizable quantum
gravity with four-derivative action and its superrenormalizable
polynomial generalizations is that the latter admit a more
sophisticated particle mass spectrum. Already in the simplest
superrenormalizable case, the theory has a six-derivative Lagrangian,
admitting either a real or complex spectrum of masses. In the case of
a real spectrum, there are the graviton, massive unphysical ghosts,
and normal particles with masses exceeding the ones of the ghosts.
It is also possible to have pairs of complex conjugate massive
ghost-like particles. We show that in both cases, these theories do
not admit a K\"all\'en-Lehmann representation and do not satisfy
the positivity criterium of consistency in terms of the fields associated
to those particles. In the main part of the work, using a 
relatively simple Euclidean scalar toy model, we show that
the theory with complex spectrum forms bound states confining
unphysical massive excitations into a normal composite particle.
Finally, we discuss the cosmological implications of such a ghost
confinement.}


\makeatletter
\gdef\@fpheader{}
\makeatother
\begin{document}
\notoc
\maketitle
\vfill

\section{Introduction}
\label{Intro}

The main problem of renormalizable models of quantum gravity is the
presence of massive unphysical ghosts and related instabilities. The
simplest  of such models is based on the fourth-derivative
covariant action{\cred. Its} particle spectrum includes {\cred a} massless 
graviton, {\cred a} tensor ghost, and a normal massive scalar particle
\cite{Stelle77}. The presence of higher derivatives and  ghosts
leads to instabilities at the classical \cite{Ostrogradski} and
quantum \cite{Veltman-63} levels and to the loss of quantum
unitarity.
Numerous attempts to resolve the contradiction between
renormalizability and unitarity by taking into account loop
corrections \cite{salstr,Tomboulis-77,antomb} were not conclusive
\cite{Johnston}. 
{The same concerns the low energy effective action of 
massless bosonic modes of (super)string theory where an infinite amount of fine-tuning
 is required  \cite{zwei} to eliminate pathological higher
derivative terms  of the effective action  \cite{ABS,ABSh2}. Although, as
remarked by Gross and Sloan  \cite{gross-sloan},  such a fine-tuning might not be necessary
 because the associated pathological ghosts would appear in the ultraviolet  regime 
where  string  perturbation theory is not reliable, which is
 not in contradiction with the unitarity of string theory \cite{Sen}.

In any case it is clear that in the field theoretical approach to quantum gravity new ideas are
required to formulate a consistent theory.
}

Extending the action of gravity by introducing additional terms with
six derivatives makes the new theory superrenormalizable
\cite{highderi} and opens some new possibilities concerning the
ghosts. For instance, in the case of a complex spectrum of ghost-like
particles, it is possible to construct a simple realization of the
proposal of \cite{salstr} and prove the unitarity of the $S$-matrix
of gravitational field \cite{Modesto-complex} in the framework of
the Lee-Wick quantization \cite{leewick,leewick2}. Still, this cannot
be a complete solution to the problem of ghosts and stability. There
is a shortcoming of an approach based on Lee-Wick applied to
gravity, which requires the conservation of energy in the closed
(isolated) system of massive particles in the
asymptotic states. In gravity, massive particles always create the
gravitational field and, in this sense, the mentioned condition is
difficult to fulfill, from the physical viewpoint.

It is clear that the main problem is the presence of ghosts and
massive normal particles in the asymptotic states. It was suggested
by Hawking in \cite{Whoisafraid} that in the four-derivative gravity,
the ghost should be considered together with a graviton. In
the six-derivative model, this proposal can be simplified by
requiring that the massive ghost forms a bound state with a normal
particle of a larger mass, or that the two complex conjugate degrees
of freedom form a bound state which is a normal (nor ghost neither
tachyon) composite particle which is harmless for stability and
unitarity.  If that mechanism works it could imply a definitive
resolution of the fundamental contradiction between renormalizability
and unitarity.

{\cred Complex-mass particles are a common feature in QCD model studies, where they are thought to signal quark and gluon confinement; one of the first suggestions in this direction was made in}~\cite{Nakanishi}. In quantum gravity, this {reasoning} of fighting massive unphysical ghosts was discussed in recent papers
\cite{Holdom,FraGhoKosh,FraGhoOka,Modesto-2023,deBrito-2023}
employing toy models inspired by those used in QCD. In the present
work, we introduce a toy model that has a propagator closely
related to one of the six-derivative quantum gravity. In this
model, we perform a basic-level quantum calculation showing that
the confinement of ghosts is possible for the sufficiently strong
coupling. The toy model of our present interest describes scalar
field with the Lagrangian possessing six derivatives. However, we
shall explore this model keeping referring to the relation with the
tensor and scalar ghosts in the six-derivative quantum gravity, as
described in the next section. A pertinent point is that since such
a theory is superrenormalizable, different from the fourth-derivative
theory, there is no gauge-fixing ambiguity in the quantum corrections
\cite{highderi,OUP}, so there are a lot of similarities in the role of
quantum contributions and one can claim the possible solution of
the ghosts problem by using the QCD analogy. In particular, in both
theories we are free to choose the model with such a spectrum of
masses which is appropriate for our purposes.

We would like to remark that {\it ghost confinement} in the present paper does not mean permanent confinement, like quark and gluon confinement in QCD, rather it is used in the sense of
complex-mass particles trapping into normal bound states. {We also use the terminology {\em ghost condensate} to mean a bound state of ghosts.}

The  remaining of the paper is organized as follows. In Sec.~\ref{sec2} we
describe the basic derivation of the propagator of quantum metric in
the six-derivative quantum gravity.  In Sec.~\ref{sec3} we discuss
the action of the toy model with a similar ghost spectrum and
reformulate its action in a useful form using auxiliary fields.  In
Sec.~\ref{sec4} the K\"all\'en-Lehmann representation for the
six-derivative model is analyzed and shown to give inconsistencies
in the UV limit.
Sec.~\ref{sec5} contains the main part of the paper, which is the
derivation of the bound states of the ghost-like fields with the
complex masses.
In Sec.~\ref{sec7} we describe the cosmological
implications of ghost confinement, including the role of the
Planck-order cut off on the cosmological perturbations and its
fundamental importance for the stability of classical solutions of
gravity. Also, we critically analyze the possibility of using the gas
of the ghost-based bound states as a dark matter candidate. Finally,
in Sec.~\ref{Concl}, we draw our conclusions.
We use  Euclidean notations in four-dimensional spacetime.

\section{Ghosts in six-derivative quantum gravity}
\label{sec2}

Let us present a brief survey of the propagator in six-derivative
quantum gravity model.  More details can be found in
\cite{OUP,QGreview}. The action of the theory has the form
\beq
S_{gen}
&=&
\int d^4 x \sqrt{g} \,
\Big\{ - \frac{1}{\ka^2} \,(R+2\La)
\,+\, \frac12\,C_{\mu\nu\al\be} \,\Phi(\cx)\,C_{\mu\nu\al\be}
\nn
\\
&&
\quad
+ \,\,
\frac12\,R \, \Th (\cx)\, R
\,+\, {\mathcal O}(R_{\dots}^3)\Big\}\,,
\label{act Phi}
\eeq
where $\ka^2 = 16\pi G = 16\pi/M_P^2$, also
$C_{\mu\nu\al\be}$ is Weyl tensor,
$\Phi(x)$ and $\Th (x)$ are linear functions of the
d'Alembert operator $\cx$ and ${\mathcal O}(R_{\dots}^3)$ stands
for the terms which are cubic in the curvature tensors (Riemann,
Ricci and scalar curvature). Independent on the choice of these terms
the theory is superrenormalizable, such that the divergences occur
only at the first three loop orders \cite{highderi}. Furthermore, the
six-derivative terms (including the terms in
${\mathcal O}(R_{\dots}^3)$ sector) are never renormalized and the
fourth-derivative terms get divergences only from the first loop,
Einstein's term only from the first and second loop and only the
cosmological constant density gets renormalized in the third loop
order.

The gauge fixing in six-derivative quantum gravity has to provide
the non-degeneracy of the highest-derivative terms in the bilinear
part of the action. This means the gauge fixing action has to be
\beq
S_{gf}
\,=\,
\frac12\,\int d^4x \sqrt{g}\,\,\,\chi_\al Y^{\al\be}\chi_\be,
\label{Sgf}
\eeq
where
\beq
&&
\chi_\al = \pa_\la h^\la_{\,\al} - \be \pa_\al h,
\nn
\\
&&
Y^{\al\be} = \big(\ga_1\eta^{\al\be}\pa^2
+ \ga_2\pa^\al \pa^\be\big)\pa^2
+ \ga_3 \pa^\al \pa^\be
+ \ga_4 \eta^{\al\be}\pa^2
+ \ga_5 \eta^{\al\be}.
\label{chiY}
\eeq
For the sake of simplicity, we use here the flat background metric
$\eta_{\al\tau}$ and the simplest parametrization of quantum metric
$g_{\al\tau}=\eta_{\al\tau} + h_{\al\tau}$. There are six arbitrary
parameters of gauge fixing $\be$ and $\ga_k$. A few more
arbitrary parameters are possible for a more general parametrization
of the quantum field. In principle, the dependence of all these
parameters is an important issue at both the tree level and in the loop
corrections. However, in the superrenormalizable gravity models
this ambiguity is much less relevant \cite{highderi,OUP,QGreview}.
For any choice of  $\be$ and $\ga_k$, the Faddeev-Popov
gauge ghosts are massless vector fields. Since these fields have
nothing to do with the massive unphysical ghosts, we shall not
consider gauge ghosts in what follows.

The quantum metric field $h_{\mu\nu}$ can be split into
irreducible parts as
\beq
&&
h_{\mu\nu} \,\,=\,\,
{\bar h}^{\bot\bot}_{\mu\nu}
+ \pa_\mu \ep^\bot_\nu + \pa_\nu \ep^\bot_\mu
+  \pa_\mu \pa_\nu \ep
+ \frac14\,h\, \eta_{\mu\nu}.
\label{h_mn}
\eeq
The tensor component (i..e., the spin-2 mode) is traceless
and transverse, i.e., ${\bar h}^{\bot\bot}_{\mu\nu} \eta^{\mu\nu}=0$
and $\pa^\mu{\bar h}^{\bot\bot}_{\mu\nu}=0$. It can be shown that
the propagator of this part of the metric does not depend on the
gauge fixing. The vector component  (i.e., the spin-1 mode) satisfies
$\pa_\mu \ep^{\bot\mu}=0$ and is gauge dependent. There are also
two scalar modes $\ep$ and $h$, one of them is gauge-dependent
while another is invariant.

Finally, there are two physical (gauge-invariant) modes, tensor and
scalar. In both cases, after the gauge fixing, the propagators include
a massless mode and two massive modes, which can be real
or complex. The propagation of the tensor mode is governed by
the Weyl-squared term and depends on the form factor $\Phi$.
The propagation of the scalar mode depends only on the square
of the scalar curvature and the function $\Th$.

It is customary to analyze the number of degrees of  freedom, ghosts
and related issues for a zero cosmological constant \cite{Stelle77}
that can be seen as working in the locally flat reference frame.
Assuming zero cosmological constant, the action for both tensor and
scalar modes $\psi = \big({\bar h}^{\bot\bot}_{\mu\nu} ,\,\,h\big)$,
on the flat background, has a general form
\beq
S_{inv} \,=\,\frac12 \int d^4x \,\,\psi
\big( \th_6 \cx^3 +  \th_4 \cx^2 +  \th_2 \cx \big) \psi\,.
\label{action-gen}
\eeq
The mass spectrum in both tensor and scalar sectors depends on the
corresponding coefficients $\th_6,\,\th_4$ and $\th_2$. In turn,
these  coefficients depend on the polynomials $\Phi(\cx)$ (tensor
sector) and $\Th(\cx)$ (scalar sector) in the action (\ref{act Phi}).
Since the situation with the two fields is similar, we can restrict
the analysis to one of them, that can be, e.g., the tensor mode.

The mass spectrum of the classical theory is defined by the poles in
the tree-level propagator.
The positions of these poles are found in the equations
\beq
p^2\Big(p^4 -  \frac{\th_4}{\th_6}\,p^2 +  \frac{\th_2}{\th_6} \Big) \,=\, 0.
\label{poles-6-geral}
\eeq
It is easy to see that there is always a massless mode (e.g., the
graviton in the tensor sector) and the two-dimensionful (massive)
solutions. For these solutions, the two physically different options
are as follows:
\vskip 1mm

{\it i)} \ \
Two real positive solutions for the poles. In this case, the
particle with a larger mass is a normal one and the lighter
massive particle is a ghost. This rule of alternating signs is
general for any polynomial gravity \cite{highderi} with the
real mass spectrum.
\vskip 1mm

{\it ii)} \ \ Complex conjugate solutions. This case is known
to provide unitary $S$-matrix in the framework of the Lee-Wick
quantization \cite{Modesto-complex,Modesto2016}. In what
follows we see that there is one more distinguishing feature of
this version of the theory - it admits, in principle, a quantum
confinement of the complex ghost-like states.
\vskip 1mm

According to the analysis of \cite{ABSh2}, if all the dimensional
parameters of the six-derivative quantum gravity are proportional to
the single massive parameter (usually assumed Planck mass), all
the masses have Planck order of magnitude. In case of the complex
spectrum this concerns the absolute values of the complex masses.
Without a special fine tuning, the real and imaginary parts of these
masses are of the same order of magnitude.

\section{Starting action in equivalent second-order form}
\label{sec3}

Consider the auxiliary fields representation of the six-derivative
action of the form
\beq
&&
\mathcal{S}_{6der}
\,=\,
\int d^4x \,\,
\frac12\, \psi \,
(- \pa^2) (- \pa^2 + m^2) (- \pa^2 + {m^*}^2) \,\psi - U(\psi).
\label{act6der}
\eeq
This action has one massless mode and two complex conjugate
modes, in the sense $m^*$ is a complex conjugate of $m$.

Let's start from the theory with three scalar fields, which has a
similar particle contents,
\beq
&&
\mathcal{S}_{3}
\,=\,
\frac{i}{2} \int d^4x \,\, \ph_1 (- \pa^2 + m^2) \ph_1
\,-\,
\frac{i}{2} \int d^4x \,\, \ph_2 (- \pa^2 + {m^*}^2) \ph_2
\nn
\\
&&
\qquad \quad
\,+\,\,
\frac{1}{2} \int d^4x \,\, \ph_3 (- \pa^2) \ph_3\,.
\label{act3fi}
\eeq
The propagators of the three fields
$\ph_{1,2,3}$, in the momentum representation are as follows:
\beq
&&
\langle \ph_1 \ph_1\rangle
\,=\, {\Big <}
\ph_1\Big(- \frac{p}{2}\Big) \ph_1\Big(\frac{p}{2}\Big)
 {\Big >}
\,=\, \frac{i}{p^2 + m^2}\,,
\nn
\\&&
\langle \ph_2 \ph_2\rangle
\,=\, {\Big <}
\ph_2\Big(- \frac{p}{2}\Big) \ph_2\Big(\frac{p}{2}\Big)
 {\Big >}
\,=\, - \,\frac{i}{p^2 + m^\ast{}^2}\,,
\nn
\\
&&
\langle \ph_3 \ph_3\rangle
\,=\, {\Big <}
\ph_3\Big(- \frac{p}{2}\Big) \ph_3\Big(\frac{p}{2}\Big) {\Big >}
\,=\, \frac{1}{p^2}\,.
\label{Greens123}
\eeq
On top of this, we assume that the three fields are independent and
therefore $\,\langle \ph_k \ph_l\rangle \sim \de_{kl}$.
Now, let us make an assumption that a certain field $\Psi$ and
$\ph_{1,2,3}$ satisfy the linear relation
\beq
\Psi \,=\, \ph_3 + \al_1 \ph_1 + \al_2 \ph_2,
\label{alphas}
\eeq
where $\al_1$ and $\al_2$ are unknown coefficients. Replacing
(\ref{alphas}) into the propagator of $\psi$, we get
\beq
&&
\langle \Psi \Psi\rangle
\,=\, {\Big <}
\Psi\Big(- \frac{p}{2}\Big) \Psi \Big(\frac{p}{2}\Big){\Big >}
\,=\,
\langle \ph_3 \ph_3\rangle
\,+\, \al_1^2 \langle \ph_1 \ph_1\rangle
\,+\, \al_2^2 \langle \ph_2 \ph_2\rangle
\nn
\\
&&
\qquad
\qquad
=\,\,
\frac{1}{p^2} - \frac{i\al_1^2}{p^2 + m^2}
- \frac{i\al_2^2}{p^2 + {m^*}^2}
\,=\,\frac{Ap^4 + Bp^2 + C}{p^2(p^2 + m^2)(p^2 + {m^*}^2)}\,.
\label{Greenpsi}
\eeq
To have an agreement with the action (\ref{act6der}), the
numerator of the last expression should be a constant. Thus,
we arrive at the  following equations for $\al_1$ and $\al_2$:
\beq
&&
A = - i\al_1^2 + i\al_2^2 + 1 = 0,
\nn
\\
&&
B = -i\al_1^2 {m^*}^2 + i\al_2^2m^2 +  {m^*}^2 + m^2 = 0,
\label{aleqs}
\eeq
while \ $C = |m|^4$ \ independent on the values of  $\al_{1,2}$.
The solutions of (\ref{aleqs}) can be easily found in the form
\beq
&&
\al_1^2 = \frac{- i{m^*}^2}{{m^*}^2 - m^2},
\qquad
\al_2^2 = \frac{im^2}{m^ 2 - {m^*}^2} = \big( \al_1^2\big)^*.
\label{alsols}
\eeq
Using these solutions, we get the propagator of the field $\Psi$
\beq
&&
\langle \Psi \Psi\rangle
\,=\,\frac{|m|^4}{p^2(p^2 + m^2)(p^2 + {m^*}^2)}\,.
\label{Green}
\eeq
This corresponds to the action
\beq
&&
\mathcal{S}_{6der}
\,=\, \frac{1}{2|m|^4}\,
\int d^4x \,\,
\Psi \,
(- \pa^2) (- \pa^2 + m^2) (- \pa^2 + {m^*}^2) \,\Psi\,, 
\label{act6psi}
\eeq
that boils down to (\ref{act6der}) after a constant reparametrization
of the field \ $\Psi = |m|^2\psi$. On the other hand, (\ref{act6psi}) can
be recognized as a particular version of the six-derivative action of
the gauge-invariant modes (\ref{action-gen}) that emerge in the model
of superrenormalizable quantum gravity with linear functions $\Phi$
and $\Th$.

In what follows, our purpose is to show that the confinement of the
massive modes $\ph_1$ and $\ph_2$ is possible for the case of a
complex mass spectrum. To arrive at this result, we will try to
simplify things as much as possible. In particular, in the analysis
of confinement we omit the massless field $\ph_3$ and also replace
the non-polynomial interactions typical for gravity, to the
particular quartic interaction of  the complex ghost-like fields
$\ph_1$ and $\ph_2$. It is clear from the relation (\ref{alphas})
that quartic interaction $\psi^4$ contains various quartic
interactions of the fields  $\ph_1$ and $\ph_2$, but we include
only some of them which are relevant for our purposes.

One of the effects of the representation (\ref{alphas}) and the last
reparametrization between $\Psi$ and $\psi$ is that the fields
$\ph_{1,2,3}$ and the six-derivative field $\psi$ have different
dimensions. In particular, the fields $\phi_1$ and  $\phi_1$  have
usual mass dimension $+1$ and the field $\psi$ has the mass
dimension $-2$.
Consequently, the coupling of the quartic interaction between the
ghosts can be dimensionless while in the six-derivative theory it is
dimensionful. A similar situation holds in the gravitational case,
where the reparametrization of the fields may change the
dimension of the coupling constants.

Let us also present an equivalent consideration based on the Gaussian
integral. As a starting action, consider the action $\mathcal{S}_{3}$
from (\ref{act3fi}) and define the common generating functional for
the three free fields,
\beq
&&
\mathcal{Z}(J)
\,=\,
\int \mathcal{D}\ph_1 \mathcal{D}\ph_2  \mathcal{D}\ph_3
\,\,\exp \Big\{
- \mathcal{S}_3 + \big(\al_1\ph_1 + \al_2\ph_2 + \ph_3\big)J
\Big\}.
\qquad
\label{Z3}
\eeq
Taking the Gaussian integrals over \ $\ph_{1,2,3}$, \ we get
\beq
&&
\mathcal{Z}(J)
\,=\,
\Big[ \Det (- \pa^2) (- \pa^2 + m^2) (- \pa^2 + {m^*}^2) \Big]^{-1/2}
\\
&&
\qquad
\times
\,\,\exp \Big\{\frac12 \int d^4x\,J \,\Big[- i\al_1^2  (- \pa^2 + m^2)^{-1}
+ i\al_2^2 (- \pa^2 + {m^*}^2)^{-1}
+ (- \pa^2)^{-1}\Big]\,J\Big\}.
\label{Z3in1}
\nn
\eeq
In the momentum representation, the expression in the square
brackets in the exponential is reproducing (\ref{Greenpsi}) and,
using the coefficients (\ref{alsols}), we arrive at
\beq
&&
\mathcal{Z}(J)
\,=\,
\Big[ \Det (- \pa^2) (- \pa^2 + m^2) (- \pa^2 + {m^*}^2) \Big]^{-1/2}
\,\exp \Big\{\frac12 \int d^4x\,J H^{-1} J\Big\}\,,
\label{Z1J}
\\
&&
\qquad
\mbox{where}
\quad
H \,=\, \frac{1}{m^2{m^*}^2}(- \pa^2 + m^2)
(- \pa^2 + {m^*}^2)(- \pa^2).
\label{tripleH}
\eeq
It is easy to recognize that (\ref{Z1J}) is a Gaussian integral
\beq
&&
\mathcal{Z}(J)
\,=\,
\int \mathcal{D}\Psi
\,\exp \Big\{- \frac12 \int d^4x\,\Psi H \Psi \,+\,\Psi J\Big\}.
\label{Z1}
\nn
\eeq
The last expression shows that we arrived at the action
(\ref{act6psi}), which is equivalent to (\ref{act6der}).
The standard interpretation of this result is that the theory
with six derivatives has either a massive ghost and a massive
normal particle \cite{highderi} or a couple of complex conjugate
ghost-like states \cite{Modesto-complex}. The question is whether
it is possible to avoid fatal instabilities in such a theory at the
classical or quantum levels. In the next section, we consider the
complex spectrum case in more detail.

\section{
Ghosts and K\"all\'en-Lehmann representation}
\label{sec4}

As we already discussed in the Introduction, the quantum gravity
theories based on higher derivative Lagrangians may be renormalizable
\cite{Stelle77}, superrenormalizable \cite{highderi,Tomboulis} or
even finite \cite{Modesto2016, mazumdar}. On the other hand,
these theories may be unstable at the classical level owing to the
Ostrogradski instabilities  \cite{Ostrogradski}. From the quantum
perspective, there are ghost fluctuations that usually lead to
instabilities \cite{Veltman-63} in the form of negative norm states,
violations of the fundamental unitary and causal properties
\cite{Stelle77,alesh,GLT,Frolov,bcmn}.

Since our present purpose is to describe the possible scheme
of confining the ghosts it is worthwhile to stress that there is no
another perspective to explain how the problem of higher derivative
ghosts can be solved. In this respect, let us add one more argument
concerning the inconsistency of the unconfined theory with higher
derivatives.
There are many ways of pointing out the conflict of higher
derivative gravity with fundamental principles of quantum field
theory. An efficient approach is showing that these theories do
not admit the   K\"all\'en-Lehmann representation
\beq
 \label{kl}
 \widehat{S}_2(p)
 \,=\,
 \!\int_0^\infty \, d\mu\, \frac{\rho(\mu)}{p^2+\mu^2},
\quad \mbox {with}
\quad
\ \rho(\mu) \geqslant  0
\eeq
of the two-point  Schwinger function $\widehat{S}_2(p^2)$ whose
existence follows immediately from the first principles. 
A useful necessary condition of  functions $\widehat{S}_{2}$ that
have a { K\"all\'en-Lehmann representation} is given by the
inequality \cite{Roberts, Widder, Krasnikov,Efimov}
\beq
\frac{ d}{ d p^2} \, p^2 S_2(p)\ > 0\,.
\label{ineq1}
\eeq
It can be shown that the theory with four- or higher-derivative
Lagrangian cannot satisfy this inequality \cite{Ezquerro}. Let us
apply the first of this condition to our
six-derivative model. In this case,
\beq
S_2(p^2) \,=\, \frac{1}{p^2 \big(p^2 + m^2\big)\big(p^2+{m^*}^2\big)}
\,\,.
\label{Wid3}
\eeq

Independent
of the magnitude of the mass $m$, in the UV we have $p^2 \gg |m|^2$.
Therefore, at high energies, $S_2(p^2) \,\approx\, 1/p^6$ and we
arrive at the estimate
\beq
\frac{d}{dp^2}\,p^2\,S_2(p^2) \,\approx\,
-\, \frac{2}{p^6}\, \leqslant\,0.
\label{Wid4}
\eeq
Thus, the inequality
(\ref{ineq1}) is violated in a theory where the propagator behaves
as $p^{-6}$ in the UV. This signals the quantum inconsistency of
the theories with six derivatives if these theories are treated in a
usual perturbative way.
The consideration presented above can be
extended to the case of real poles and other higher derivative
models, where there is a generalization of   K\"all\'en-Lehmann
representation \cite{gclr}.
This extension will be discussed elsewhere, while here
we consider only the six-derivative models.

Different from the UV, quantum gravity models with higher derivatives
may be consistent in the IR, where the higher derivative terms are
regarded as small corrections to GR by definition
\cite{Simon-90,ParSim} or if there are imposed restrictions on the
frequencies of the quantum perturbations \cite{HD-Stab,PP} by
imposing a Planck-order cut-off.  The pertinent question is what can
be the origin of such a UV cut-off. In the next section, we argue
that the corresponding mechanism could be a confinement of complex
massive ghost-like states that emerge in the six or more-derivative
quantum gravity with the complex mass spectrum.

\section{Bound states with ghost-like complex massive particles}
\label{sec5}

Consider the following six-derivative Euclidean
Lagrangian for the field $\psi$
\beq
\mathcal{L} = \frac12\, \psi \big[ -\partial^2 \big(-\partial^2 + m^2\big)
\big(-\partial^2 + m^{*2}\big) \big] \psi - U(\psi),
\label{act6der-II}
\eeq
where $m^2 = m^2_R + i m^2_I$ is a complex squared mass and
$U(\psi)$ is for now unspecified potential. This potential is the
main difference to the free model (\ref{act6der}) discussed above.

The momentum-space
free propagator $D_\psi(p)$ of the $\psi$-field is given by:
\beq
D_\psi(p) = \frac{1}{p^2 \left(p^2 + m^2\right) \left(p^2 + m^{*2}\right)}\,.
\label{Dpsi}
\eeq
Using the results of Sec.~\ref{sec3},
this propagator can be split as a sum of three propagators:
\beq
D_\psi(p) = \frac{1}{|m|^4} \frac{1}{p^2}
- \frac{iA}{p^2 + m^2}
+ \frac{iA^*}{p^2 + m^{*2}}
\hspace{0.5cm}\text{with}\hspace{0.5cm}
A = \frac{1}{|m|^4} \frac{m^2}{2 m^2_I}\,.
\label{Dpsi-sum}
\eeq
As we know, the Lagrangian in \eqref{act6der-II} is equivalent
to the following one:
\beq
\mathcal{L} = \frac12 \phi \big(- \partial^2 \big) \phi
+ \mathcal{L}_{\rm gh} ,
\eeq
where the first term leads to the first term in \eqref{Dpsi-sum}
(after absorbing the $|m|^4$ in the field
$\phi$) and $\mathcal{L}_{\rm gh}$ is the Lagrangian of two ghost
fields $\varphi_1$ and $\varphi_2$ (that absorb the $A$ and $A^*$)
that leads to the other two propagators in~\eqref{Dpsi-sum}:
\beq
\mathcal{L}_{\rm gh}
= \frac12 \varphi_1\big[ i \, \big(-\partial^2 + m^2\big)\big]
\varphi_1
+ \frac12 \varphi_2\big[  - i \, \big(-\partial^2 + m^{*2}\big)\big]
\varphi_2
- U(\varphi_1,\varphi_2)\,,
\label{act2gh}
\eeq
where $U(\varphi_1,\varphi_2)$ is a Hermitian potential that in
principle is related to the $U(\psi)$ potential. The fact that
$\mathcal{L}_{\rm gh}$ leads to a complex Euclidean
action\footnote{A complex Euclidean action occurs e.g. in QCD,
when one needs to consider a baryon chemical potential~$\mu_B$
to treat high-density nuclear and quark matter. The chemical
potential contributes with a complex term to the fermionic part
of the QCD action, namely $- i \int d^4x\, {\Psi}^\dag  \mu_B  \Psi$,
where $\Psi$ is the quark field. A complex Euclidean action leads
to what is known as the {\em sign problem}; it obstructs the use of
the Monte Carlo method, which is the basis of lattice QCD
simulations. This obstruction is the main reason for the lack of
progress in the knowledge on the properties of the matter in the
interior of neutron stars.} should not be of concern since the
generating functional
\beq
Z_{\rm gh} = \int \mathcal{D}\varphi_1 \mathcal{D}\varphi_2
\, e^{-S_{\rm gh}} \hspace{0.5cm}\text{where}
\hspace{0.5cm}S_{\rm gh} = \int d^4x \, \mathcal{L}_{\rm gh}
\eeq
is real because $\varphi_1$ and $\varphi_2$ are dummy integration variables.

Let us start with the survey of the dimensions of field and couplings.
As we mentioned above, the dimensions of the fields get modified
when using auxiliary fields. Consider this in detail. The field $\psi$
in \eqref{act6der-II} has mass-dimension
$[\psi] = - 1$.  Consider a possible interaction term in $U(\psi)$
\begin{equation}
U(\psi) =\frac{ \lambda}{4 !} \, \psi^4,
\qquad
\mbox{hence}
\qquad
[\lambda] = 8,
\label{ex-U}
\end{equation}
wheras the dimensions of the fields $\varphi_i$  are
$[\varphi_i] = 1$ 

To investigate the possibility that the ghost fields $\varphi_1$ and
$\varphi_2$ form two-particle bound states, we make (perhaps) the
simplest possible computation. Let $O_{\varphi_1\varphi_2}(x)$ be
the {\cred scalar} composite operator (correlator)
\beq
O_{\varphi_1\varphi_2}(x) = \varphi_1(x) \varphi_2(x).
\eeq
Then, we consider the correlation function $G(x,y)$
\begin{align}
C(x,y) &= C(x-y) = \langle O_{\varphi_1\varphi_2}(x) \,
O_{\varphi_1\varphi_2}(y) \rangle \nn \\[0.25cm]
&= \frac{1}{Z_{\rm gh}} \int \mathcal{D}\varphi_1 \mathcal{D}\varphi_2 \,
O_{\varphi_1\varphi_2}(x) \,
O_{\varphi_1\varphi_2}(y) \; e^{-S_{\rm gh}}.
\end{align}
To compute $C(x-y)$, we need to specify  $U(\varphi_1,\varphi_2)$.
We proceed within an effective field theory (EFT) perspective, i.e., consider all possible interactions with coefficients to be adjusted phenomenologically. The lowest-order terms are the renormalizable ones, namely
\beq
U(\varphi_1,\varphi_2)
\,=\,
\frac{1}{4!} \lambda \varphi^4_1
+ \frac{1}{4!} \lambda \varphi^4_2
+ \frac{1}{4} \lambda_{12} \varphi^2_1 \varphi^2_2.
\eeq
where $\lambda_{12}=\lambda/3 $.  
We then compute $C$ in perturbation theory; the $\mathcal{O}(\lambda_{12})$
one-loop  contribution, $C^{\rm 1-loop}$, is given by
\beq
&&
C^{\rm 1-loop}(x,y)
\,=\,
D_{\varphi_1}(x-y) D_{\varphi_2}(x-y)
\nn
\\
&&
\qquad\qquad
+ \,\, \lambda_{12}  \int d^4z \,
D_{\varphi_1}(x-z) D_{\varphi_2}(x-z) D_{\varphi_1}(z-y)
D_{\varphi_2}(z-y),
\label{C1}
\eeq
where $D_{\varphi_1}$ and $D_{\varphi_1}$ are the ghost-field
free propagators. In momentum space, they are given by
\beq
D_{\varphi_i}(x-y) \,=\,
\int \frac{d^4p}{(2\pi)^4}\,\, e^{- i p\cdot(x-y)}\,\,
D_{\varphi_i}(p), \hspace{1.0cm}i=1,2\,,
\label{Dphi-FT}
\eeq
where
\beq
D_{\varphi_1}(p)
= \frac{i}{p^2 + m^2} \hspace{1.0cm}
\text{and}\hspace{1.0cm}
D_{\varphi_2}(p) = \frac{-\,i\,\,}{p^2 + m^{*2}}\,.
\label{Dphi12}
\eeq
Note that the factors $i$ and $-i$ in these propagators come from
the $i$ and $-i$ multiplying respectively $\big(-\partial^2 + m^{2}\big)$
and $\big(-\partial^2 + m^{*2}\big)$ in \eqref{act2gh}.

Consider the first term in \eqref{C1}:
\begin{align}
D_{\varphi_1}(x-y) D_{\varphi_2}(x-y) &= \int \frac{d^4k'}{(2\pi)^4}
\frac{d^4k}{(2\pi)^4} \, e^{-i k'\cdot(x-y) - i k\cdot(x-y)} \,
D_{\varphi_1}(k') \, D_{\varphi_2}(k) \nn \\[0.25true cm]
&= \int \frac{d^4p}{(2\pi)^4} \, e^{- i p \cdot (x -y) }
\int \frac{d^4k}{(2\pi)^4} \, D_{\varphi_1}(p-k) \, D_{\varphi_2}(k)
\nn \\[0.25true cm]
&= \int \frac{d^4p}{(2\pi)^4} \, e^{- i p \cdot (x -y) } \; G_B(p)\,,
\label{DD}
\end{align}
where $G_B(p)$ is the {\em bubble integral}
\beq
G_B(p)
\,= \,\int \frac{d^4k}{(2\pi)^4} \, D_{\varphi_1}(p-k)
\, D_{\varphi_2}(k) .
\label{bubble}
\eeq
Using the result \eqref{DD} twice and integrating over~$z$,
one can show that the second term in \eqref{C1} can be written as:
\beq
&&
\int d^4z \,
D_{\varphi_1}(x-z) D_{\varphi_2}(x-z) D_{\varphi_1}(z-y) D_{\varphi_2}(z-y)
\nn
\\
&&
\qquad
= \,\,
\int \frac{d^4p}{(2\pi)^4} \, e^{- i p \cdot (x -y) } \; G_B(p) \; G_B(p).
\label{4Ds}
\eeq
Next, defining the Fourier transform $C(p)$ of $C(x-y)$ through
\beq
C(x -y) = \int \frac{d^4p}{(2\pi)^4}\, e^{- i p\cdot(x-y)}\; C(p),
\eeq
one can then write \eqref{C1} in momentum space
as\footnote{To lighten the notation we omit the index
${\rm 1-loop}$ in $C^{\rm 1-loop}(p)$.}
\beq
C(p) = G_B(p) + G_B(p) \big[ \lambda_{12} \, G_B(p) \big].
\eeq

One can now iterate this one-loop result to obtain a Dyson's
type of equation,
\begin{align}
C(p) &
\,=\, G_B(p) + G_B(p) \big[ \lambda_{12} \, G_B(p) \big]
+ G_B(p) \big[ \lambda_{12} \, G_B(p) \, \lambda_{12} \, G_B(p) \big]
\nn
\\[0.25true cm]
&\;\;\;\,+\, G_B(p) \big[  \lambda_{12} \, G_B(p) \, \lambda_{12} \, G_B(p) \,
\lambda_{12} \, G_B(p)\big] + \cdots  \nn\\[0.25true cm]
&
\quad
= \,\,G_B(p) \, \sum^\infty_{n=0} \big[\lambda_{12}\,G_B(p)\big]^n
\,\,=\,\, \frac{G_B(p)}{1 - \lambda_{12}\,G_B(p)}\,\,.
\label{GB-pole}
\end{align}

There is a physical bound state if $C(p)$ has a pole at a value
$p^2 = - {\cal M}^2$ and the residue at the pole is positive. There
will be a pole at $p^2 = - \mathcal{M}^2$ when
\beq
1 - \lambda_{12}\,G_B(p){\Big |}_{p^2 = - \mathcal{M}^2} = 0.
\label{pole-cond}
\eeq
If there is a pole, there will exist a critical value of the coupling
$\lambda_{12}$ where the pole first appears. Let us compute the
one-loop integral by using spherical coordinates and assuming that
we have a pair of complex masses $m^2=(1+i)\mu^2$ and
$m^{\ast}{}^2=(1-i)\mu^2$. We get
\beq
&&
G_B(p) \,=\,
\frac{1}{ 4 (2\pi)^2\,p^2}\! \int_0^\infty\!\! dk\, k\,
\Big\{ (1-i)\mu^2 + k ^2\Big\}^{-1}\,\,
\Big\{(1+i)\mu^2 + k^2 +  p^2\!
\nn
\\
&&
\qquad \qquad \,\, - \,\,\,
\sqrt{\big[(1+i)\mu^2 + (k + p)^2\big]\, \big[(1+i)\mu^2 + (k - p)^2)\big]}
\Big\}.
\qquad
\label{formerred}
\eeq
The last integral is logarithmically divergent, but this divergence
can be easily renormalized by momentum substraction
\begin{equation}
G_B^R(p) \,=\, G_B(p) \,-\, G_B(p_0).
\end{equation}
The result  for \ $1- \lambda_{12}G_B^R(i p)$ \ is displayed in Figure~1
once the substation point has been fixed at $p_0^2=1$ with unit mass
scale  $\mu=1$ and coupling constant $\lambda_{12}= \pi^5$. The  presence of
a pole at ${\cal M}^2 =1.56$  supports the existence of a bound ghost
state under these conditions.

\begin{figure}[htb]
\begin{center}
$\,$\includegraphics[angle=0,width=11cm] {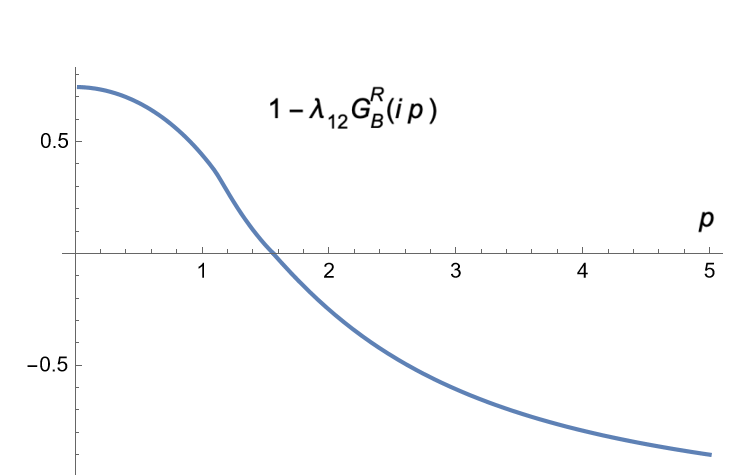}
$\,$
\end{center}
\begin{quotation}
\caption{The denominator in  the right hand side of Eq.~\eqref{GB-pole} 
as a function of \ $-i \mathcal{M} =  \,p$, where
we have chosen $\lambda_{12}= \pi^5$ and ${\mu}^2=1$.}
\end{quotation}
\label{fig2}
\end{figure}

The mass of the bound state $\mathcal{M}$ is the solution of
(\ref{pole-cond}). As mentioned, to be a physical pole, its
residue should be positive. To obtain the residue, we expand
$C(p)$ around the point $p^2 = - \mathcal{M}^2$ to get
\beq
&&
C(p)\Big{|}_{p^2 \approx - \mathcal{M}^2}
\,=\,
\frac{ \,G_B^R( -\mathcal{M}^2) + \cdots }{ 1
- \lambda_{12}\,G_B^R( -\mathcal{M}^2) - (p^2 + \mathcal{M}^2)
\,\lambda_{12} \, \,G_B^R{}^\prime( -\mathcal{M}^2) + \cdots}
\nn
\\
&&
\qquad
\qquad
\qquad
\qquad
=\, \frac{R_G^R}{p^2 + \mathcal{M}^2} + \cdots\,\,,
\quad
\label{Cp}
\eeq
with the residue $R_G^R$ given by
\beq
R_G^R = - \frac{1}{\lambda_{12}^2\,G_B^R{}^\prime( -\mathcal{M}^2) }\,.
\hspace{0.5cm}
\label{RG-def}
\eeq
For a positive $R_G$, one must have
$\,G_B^R{}^\prime( -\mathcal{M}^2) < 0$.

Now, from the behavior of the two-point function displayed in Figure
1 it follows that the residue at the pole $\mathcal{M} = 1.56$ is
positive  $R_G > 0$. Thus, it  {corresponds} to a physical pole and marks
the appearance of a ghosts condensate.

The existence of a physical solution with a bound condensate of
ghosts depends on the value of the complex masses of the pair of
ghosts $(1\pm i)\mu$, the coupling constant of the model $\lambda_{12}$ and
the renormalization momentum substraction point  $p_0$. But in
our case, we have shown that  that there is a window of coupling
constants where the condensation of ghosts is possible. The
critical values of the coupling constant $\lambda_{12}^\pm$ enclosing such a
window are 
$\lambda_{12}^-=0.68 \pi^5$  and $\lambda_{12}^+=3.91 \pi^5$, 
i.e. only if $\lambda_{12}^-\le \lambda_{12} \le \lambda_{12}^+$
the condensation occurs.

Our analysis indicates that there is a possibility to form bound
states from the pair of complex conjugate massive ghosts. Similar results were obtained when using the $\overline{\rm MS}$ renormalization scheme.
We skip the details of these {\cred calculations,} since the results are
equivalent.

It is important to {note} that the possibility of having a second pole with a negative residue, which {would} correspond to a composite {\cred ghost,} can be discarded because {\cred the function $1- \lambda_{12} 
G_B^R(-p^2)$, as can be seen in Figs.~1~and~2, is a monotonically decreasing function of $p$, reaching an asymptotic value in the $p^2\to \infty$ limit: 
\begin{equation} 
1- \lambda_{12} G_B^R(-p^2) = -1.18 + {\mathcal O}(1/p^2),
\end{equation}
which is almost saturated at maximum momentum displayed in Fig.~2.
}

\begin{figure}[htb]
\begin{center}
$\,$\includegraphics[angle=0,width=11cm] {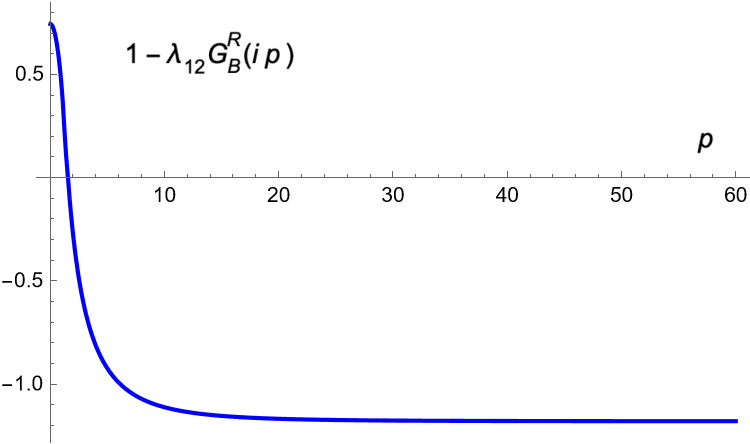}

$\,$
\end{center}
\begin{quotation}
\caption{The analytic continuation to imaginary values of $p$ of denominator of the right hand side of {Eq.~\eqref{GB-pole}. Here,}
we have chosen $\lambda_{12}= \pi^5$ and ${\mu}^2=1$.}
\end{quotation}
\label{fig2}
\end{figure}

{\cred One can further substantiate the possibility of forming bound
states out of a pair of complex conjugate massive ghosts by examining an alternative scalar composite field operator, for example: }
\beq
\widetilde{O}_{\varphi_1\varphi_2}(x) = \varphi_1(x)^2+ \varphi_2(x)^2.
\eeq
{\cred This operator has the same quantum numbers of $O_{\varphi_1\varphi_2}$ and, as such, is equally suited to study the existence of bound states, as would any other composite operator with the same quantum numbers~\cite{Zimmermann:1987ne}. An explicit calculation, at the same level of approximation as in the previous calculation, shows that the correlation function 
\begin{equation} 
\widetilde{C}_B(x,y)=\langle \widetilde{O}_{\varphi_1\varphi_2} (x) \widetilde{O}_{\varphi_1\varphi_2} (y) \rangle,
\end{equation} 
also} shows the presence of a single real pole in the corresponding  renormalized correlator{\cred :
\begin{equation}
\widetilde{C}{}_B^R(p^2)= \,\, \frac{\widetilde{G}{}^R_B(p)}{1 - \frac{\lambda}{2} \,\widetilde{G}{}^R_B(p)},
\label{GBB-nopole}
\end{equation}
where} ${\widetilde{G}{}^R_B(p)}={\widetilde{G}{}_B(p)}-{\widetilde{G}{}_B(1)}$
with ${\widetilde{G}{}_B(p)}={\widetilde{G}{}^+_B(p)}+{\widetilde{G}{}^-_B(p)}$ and
\beq
&&
\widetilde{G}^\pm_B(p) \,=\,
-\frac{1}{ 4 (2\pi)^2\,p^2}\! \int_0^\infty\!\! dk\, k\,
\Big\{ (1\pm i)\mu^2 + k ^2\Big\}^{-1}\,\,
\Big\{(1\pm i)\mu^2 + k^2 +  p^2\!
\nn
\\
&&
\qquad \qquad \,\, - \,\,\,
\sqrt{\big[(1\pm i)\mu^2 + (k + p)^2\big]\, \big[(1\pm i)\mu^2 + (k - p)^2)\big]}
\Big\}.
\qquad
\label{formerred2}
\eeq
Indeed,  the denominator $1- \lambda \widetilde{G}{}_B^R(i p)$ {\cred has a single zero at ${\cal M} =5.92$ {\cred with a positive residue $R_{\widetilde{G}} > 0$, as one can see in Figure~3}.  The window of coupling
constants $\lambda$ for the occurrence of the bound state is unbounded,
$0<\lambda<\infty$.}
This is due to the asymptotic logarithm behavior of 
{\cred $\widetilde{G}{}_B^R(i p) = 0.031 + ({4}/{\pi^2} + \mathcal{O}(1/p^2))\log p$} in  the $p^2\to \infty$ limit. {\cred This different coupling dependence from the previous example is expected, as both the composite operators and the coupling are scale dependent (they run with the renormaliation scale), but the pole position should be scale independent. The study of such features is out of the scope of this initial study and is reserved for a future publication.}
\begin{figure}[htb]
\begin{center}
$\,$\includegraphics[angle=0,width=11cm] {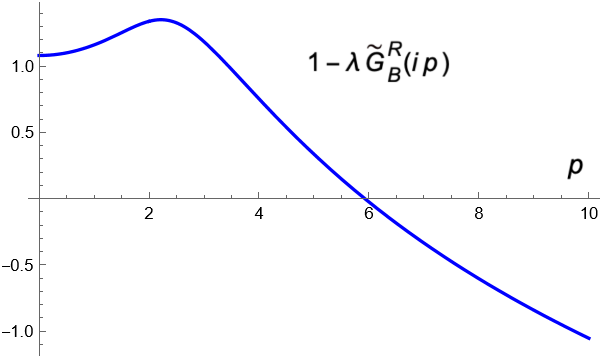}
$\,$
\end{center}
\begin{quotation}
\caption{$1-\lambda \widetilde{G}{}^R_B(i p)$ as a function of $p$, where
we have chosen ${\mu}^2=1$ and $\lambda=\pi^5/4$.}
\end{quotation}
\label{fig3}
\end{figure}

The above results give a real hope to observe a similar effect in the
superrenormalizable quantum gravity model with six derivatives,
and probably also in the cases of $2+2n$ derivatives, where
$n=3,4,...$, because these models admit mass spectrum consisting
from complex conjugate pairs of poles plus the graviton. 
\section{Cosmological implications of ghost confinement}
\label{sec7}

It would be interesting to extend the scheme of confinement of
ghosts to the real quantum gravity and arrive at the definite answer
concerning the bound states, by using the methods borrowed from
nonperturbative QCD. However, this will not be an easy task because
even in QCD, definite analytic results concerning confinement of
quarks are difficult to obtain. On the other hand, it is interesting
to speculate about possible observable manifestations of physical
bound states of complex mass ghosts. Since the mass of such a
composite particle and the corresponding cutoff on the energy of
the original gravitational modes are of the Planck order of
magnitude, such observations can belong only to early cosmology,
where we meet a growing amount and quality of available data.

Is it possible to find cosmological traces of the confinement of the
ghost-like complex modes? Thinking about particle physics and
accelerator experiments, the answer is
negative because the masses of the bound states are of the Planck
order of magnitude, far beyond the scale of the present or future
particle physics. An additional feature is that the composite
particles do not bring any charges except the mass and, therefore,
do not engage in any interactions except gravitational.

The situation is different in the case of early cosmology or black
hole perturbations of the metric or (in the cosmological case)
density. The first effect of ghost confinement is imposing a
Planck cut-off on the energy of the gravitational perturbations. For
the sake of definiteness, consider the early cosmology. Looking
back in time, as far as a cosmic perturbation becomes
trans-Planckian, the pair of complex conjugate ghosts is created
and gets confined into a bound state. This situation rules out the
observation of the trans-Planckian physics in both cosmology
(see e.g., \cite{Martin-2001,Starobinsky-2001}) and black hole
evaporation (see, e.g., \cite{Barbado-2011}). The Planck-order
cut-off in the case of cosmological perturbations may be, in
principle, detected by future observational facilities.

It is important to note that the same happens if the gravitons
concentrate in a locality with a Planck-order density of
energy. Such a concentration is a necessary element of creating
a pair of massive complex ghosts from the vacuum forming
the bound states rule out the instabilities created by massive
ghosts, as discussed in \cite{HD-Stab,PP}. All in all, the definite
answer about forming the bound state of complex ghosts would
resolve the contradiction between renormalizability and unitarity
in quantum gravity.

Finally, looking forward in time from the Planck-energy epoch in
cosmology, what could be the possible effect of the bound states in
the later epochs? For example, are these particles, with the masses
of the Planck order of magnitude, which interact only gravitationally,
realistic candidates to be Dark Matter? To address this question,
let us make a numerical estimate. The first thing to note is that the
complex ghosts bound states are created at the Planck energy scale,
i.e., at energies much higher than the ones typical for inflation.

As in the rest of this paper, we follow the simplest possible
approach. At the moment of creating the bound states, we assume
that the energy scale is ${\mathcal E}_{in} = M_P$ and the initial
energy density of these composite particles is
$\rho^{BS}_{in}(M_P) = M_P^4$.
Consider that right after this point the inflation starts. This means
that we are making a too optimistic estimate because the bound states
could be dissolved between the Planck epoch and the beginning of
inflation. However, since this point does not change the result, we
can use it as a simplification.
Now, we evaluate the evolution of the density of bound states at the
energy ${\mathcal E} \sim 1/a(t)$, corresponding to the conformal
factor $a(t)$ of the cosmological expansion. The described conditions
produce the result
\beq
\rho^{BS}({\mathcal E})
\,\,\propto \,\, \rho^{BS}_{in} \,
\Big(\frac{{\mathcal E}}{{\mathcal E}_{in}}\Big)^3
\,\,=\,\, M_P^4 \Big(\frac{a_{in}}{a}\Big)^3.
\label{rhoBS}
\eeq
For the density of bound states at the end of $N$ e-folds
inflation, we get
\beq
\rho^{BS}({\mathcal E}_{end})
\,\,\propto \,\,
M_P^4 \, e^{-3N}.
\label{rhoBS-N}
\eeq
For the critical density at the same epoch, one can use the Friedmann
equation
\beq
\rho_c ({\mathcal E}_{end})
\,=\, \frac{3}{8\pi G}\,H_{end}^2
\,=\, \frac{3M_P^2}{8\pi}\,H_{end}^2.
\label{FriEq}
\eeq
Using the standard estimate $H_{end} \approx 10^{12}\,{\rm GeV}$ for the
Hubble parameter at the end of inflation, taking the number of e-folds
$N=70$ and $M_p = 10^{19}\,{\rm GeV}$, we get
\beq
\rho^{BS}({\mathcal E}_{end})
\,\,\propto \,\,
\rho_c ({\mathcal E}_{end})\,\times\, 10^{-78}\,.
\label{NumerEvo}
\eeq
After the end of inflation, the density of bound states continues 
dissolving, however in the early Universe, the rest of the matter may 
show faster decrease of critical density. Then, the ratio with the 
critical density may loose up to 25 orders of magnitude, but the 
hierarchy between this density and density of the ghost's bound 
states still remains very strong. 

This result implies that there is no chance to regard the gas of the
Planck mass-bound states of complex ghosts as a Dark Matter
candidate. The point is that this gas is generated too early and
expands too much after the Planck-energy epoch.

\section{Conclusions}
\label{Concl}

We have analyzed the toy model with a pair of complex conjugate
massive unphysical ghost-like states, i.e., the mass spectrum which
is typical for the six-derivative superrenormalizable quantum gravity.
It was shown that this theory may describe the confinement of the
ghost states if the coupling constant is large enough. This condition
can be fulfilled in the models of quantum gravity by either choosing
appropriate parameters in the action (\ref{act Phi}) or by describing
the force between the two ghosts by exchanging gravitons.

The bound states of the massive complex ghost-like particles form
the Planck-density gas in the early Universe. The subsequent
expansion between the Planck scale and the end of inflation strongly
dissolves this gas, such that it cannot be even a small part of the
Dark Matter. Thus, the unique potentially observable consequence of
the ghost confinement is the cut-off of the Planck order of magnitude
to the energy density of gravitons. The two relevant aspects of this
quantum-origin cut-off are the non-existence of the trans-Planckian
effects on cosmic perturbations \cite{Martin-2001,Starobinsky-2001}
and, perhaps the most important, the explanation of why there are no
Planck-order frequencies in the initial seed of cosmic perturbations.
The last issue is a cornerstone in explaining the stability of
classical cosmological solutions in the presence of higher
derivatives \cite{HD-Stab,PP}.  We note that there is a significant
interest in discussing other physical outputs of a Planck-scale
cut-off on the energy density of gravitons (see, e.g., \cite{Dvali}).
The confinement of ghosts which we found in the toy model of
superrenormalizable quantum gravity, may offer a solid background
to such a cut-off.

Looking beyond the scope of the present work, further explanation
of the subject requires elaborating models {to quantum gravity and, on the other hand, 
{making more complete analyses} of the formation of bound states. We hope to advance in these directions in  future works. 

{\cred In particular, there} are two aspects of the described bound states that were not discussed {\cred in the present paper but will} be reported in a separate future work. The first one concerns the existence of the K\"all\'en-Lehmann
representation for the bound states. This representation is possible
and can be demonstrated in ways similar to what was done in
QCD \cite{Sorella1,Sorella2,Sorella3}. {\cred Since these proofs
are technically elaborated, we leave such an analysis for a separate study}. The
second issue concerns the six-derivative theory with a real mass
spectrum. As  mentioned above, such a theory has a massive
ghost and a normal particle with a mass exceeding the mass of the
ghost. The preliminary answer {regarding bound states in this case is negative,} but it was obtained with the same model of
self-interaction between ghosts that was used in the complex
case. {\cred We also leave this issue for future work}.

\section*{Acknowledgements}

The authors are grateful to ICTP-SAIFR for organizing our event
``Minicourse on perturbative and nonperturbative treatment of
quantum gravity problems'', during which much of this work
was done.
This work was supported by Spanish Grants No.
PGC2022-126078NB-C21 funded by MCIN/AEI/10.13039/501100011033,
Diputaci\'on General de Arag\'on-Fondo Social Europeo (DGA-FSE)
Grant No. 2020-E21-17R of the Aragon Government; and the European Union,
NextGenerationEU Recovery and Resilience Program on {\em Astrof\'{\i}sica y
F\'{\i}sica de Altas Energ\'{\i}as},
CEFCA-CAPA-ITAINNOVA (M.A); {\it Programa Investigo} 
funded by the European Union - Next Generation EU - Plan de Recuperaci\'on, 
Transformaci\'on y Resiliencia (M.P.);
Conselho Nacional de Desenvolvimento
Cient\'{i}fico e Tecnol\'{o}gico - CNPq for the partial support
under the grants 305122/2023-1 (I.Sh.), 305894/2009-9 and
305000/2023-3 (G.K.); Funda\c{c}\~{a}o de Amparo \`{a}
Pesquisa do Estado de S\~{a}o Paulo (FAPESP), grant No.
2013/01907-0 (G.K.).


\begin{thebibliography}{99}
\renewcommand{\baselinestretch}{1.1}

\bibitem{Stelle77} K.S. Stelle,
{ \it Renormalization of higher derivative quantum gravity},
Phys. Rev. {\bf D16} (1977) 953.

\bibitem{Ostrogradski} M.V.~Ostrogradski,
\textit{Memoires sur les equations differentielles relatives
au probleme des isoperimetretres,}
Mem. Acad. St. Petersburg \textbf{6} (1850) 385. 

\bibitem{Veltman-63} M.J.G. Veltman,
{\it Unitarity and causality in a renormalizable field theory
with unstable particles},
Physica {\bf 29} (1963) 186. 

\bibitem{Tomboulis-77} E.~Tomboulis,
{\it $1/N$ expansion and renormalization in quantum gravity},
Phys. Lett.  {\bf B70} (1977)  361;
{\it Unitarity in Higher derivative quantum gravity},
Phys. Rev. Lett. {\bf 52}  (1984)  1173.

\bibitem{salstr}  A.~Salam and J.~Strathdee,
{\it Remarks on high-energy stability and renormalizability
of gravity theory},
Phys. Rev.  {\bf D18} (1978)  4480.

\bibitem{antomb} I.~Antoniadis and E.T.~Tomboulis,
{\it Gauge invariance and unitarity in higher derivative
quantum gravity},
Phys. Rev. {\bf D33} (1986)  2756.

\bibitem{Johnston} D.A.~Johnston,
{\it Sedentary ghost poles in higher derivative gravity,}
Nucl. Phys. {\bf B297} (1988) 721.

\bibitem{zwei}
B. Zwiebach, {\it Curvature squared terms and string theories},
Phys. Lett. {\bf B156} (1985) 315.

\bibitem{ABS} A. Accioly, B.L. Giacchini and I.L. Shapiro,
{\it Low-energy effects in a higher-derivative gravity model
with real and complex massive poles},
Phys. Rev. D {\bf 96}, 104004 (2017),
arXiv:1610.05260.

\bibitem{ABSh2} A.~Accioly, B.L.~Giacchini and I.L.~Shapiro,
{\it On the gravitational seesaw in higher-derivative gravity},
Eur. Phys. J.  {\bf C77} (2017) 540,
gr-qc/1604.07348.

\bibitem {gross-sloan}  D. J. Gross and J. H. Sloan,
{\it The Quartic Effective Action for fhe Heterotic String}. 
Nucl. Phys.   {\bf B291} (1987) 41.

\bibitem {Sen}  A  Sen,
{\it Unitarity of superstring field theory}. 
JHEP. {\bf 2016} 115 (2016).

\bibitem{highderi} M. Asorey, J.L. L\'opez and I.L. Shapiro,
{\it Some remarks on high derivative quantum gravity,}
Int. Journ. Mod. Phys. {\bf A12} (1997) 5711,
hep-th/9610006.

\bibitem{Modesto-complex} L.~Modesto and I.L.~Shapiro,
{\it Superrenormalizable quantum gravity with complex ghosts,}
Phys. Lett. {\bf B755} (2016) 279,
arXiv:1512.07600.

\bibitem{leewick} T.D. Lee and G.C. Wick,
\textit{Negative metric and the unitarity of the $S$ matrix,}
Nucl. Phys. {\bf B9} (1969) 209. 

\bibitem{leewick2} T.D. Lee and G.C. Wick,
\textit{Finite Theory of Quantum Electrodynamics,}
Phys. Rev. {\bf D2} (1970) 1033. 

\bibitem{Whoisafraid} S.W. Hawking,
\textit{Who's afraid of (higher derivative) ghosts?}
(Preprint-86-0124, Cambridge, 1985).

\bibitem{Nakanishi} N.~Nakanishi,
\textit{A possible field-theoretical model of quark confinement,}
Prog. Theor. Phys. \textbf{54} (1975) 1213.

\bibitem{Holdom} B. Holdom and J. Ren,
\textit{QCD analogy for quantum gravity,}
Phys. Rev. \textbf{D93} (2016) 124030,
arXiv:1512.05305.

\bibitem{FraGhoOka} M. Frasca, A. Ghoshal and N. Okada,
\textit{Confinement and renormalization group equations in
string-inspired nonlocal gauge theories,}
Phys. Rev. \textbf{D104} (2021) 096010,
arXiv:2106.07629.

\bibitem{FraGhoKosh} M. Frasca, A. Ghoshal and A.S. Koshelev,
\textit{Confining the complex ghosts out,}
Phys. Lett. \textbf{B841} (2023) 137924,
arXiv:2207.06394.

\bibitem{Modesto-2023} J. Liu, L. Modesto and G. Calcagni,
\textit{Quantum field theory with ghost pairs,}
JHEP \textbf{02} (2023) 140,
arXiv:2208.13536.

\bibitem{deBrito-2023} G.~P.~de Brito,
\textit{Quadratic gravity in analogy to quantum chromodynamics:
Light fermions in its landscape,}
Phys. Rev. \textbf{D109} (2024)  086005,
arXiv:2309.03838.

\bibitem{OUP} I.L. Buchbinder and I.L. Shapiro,
\textit{ Introduction to Quantum Field Theory with Applications to
Quantum Gravity,} (Oxford University Press, 2021).

\bibitem{QGreview} I.L.~Shapiro,
\textit{The background information about perturbative quantum
gravity,}
Chapter in. {\sl Handbook of Quantum Gravity}
(Springer Singapore, 2023),
arXiv:2210.12319.

\bibitem{Modesto2016} L.~Modesto,
\textit{Super-renormalizable or finite Lee-Wick quantum gravity,}
Nucl. Phys. \textbf{B909} (2016) 584, 
arXiv:1602.02421.

\bibitem{Tomboulis} E.T. Tomboulis,
{\it Nonlocal and quasilocal field theories,}
Phys. Rev. {\bf D92} (2015) 125037,
arXiv:1507.00981,
{\it Superrenormalizable gauge and gravitational theories,}
hep-th/9702146.

\bibitem{mazumdar}
T. Biswas, A. Conroy, A.S. Koshelev, A. Mazumdar,
\textit{Generalized ghost-free quadratic curvature gravity,}
Class. Quant. Grav. \textbf{31} (2014) 015022
[Erratum: Class. Quant. Grav. \textbf{31} (2014) 159501],
arXiv:1308.2319.

\bibitem{alesh} M. Asorey, L. Rachwal, I.L. Shapiro,
\textit{Unitary issues in some higher derivative field theories,}
Galaxies \textbf{6} (2018) 1, 23,
arXiv:1802.01036.

\bibitem{GLT} D.M. Gitman and I.V. Tyutin,
{\it Quantization of Fields with Constraints,}
(Springer Berlin-Heidelberg, 1990).

\bibitem{Frolov} V.P. Frolov and A. Zelnikov,	
\textit{Radiation from an emitter in the ghost free scalar theory,}
Phys. Rev. \textbf{D93} (2016) 105048,
arXiv:1603.00826.

\bibitem{bcmn} F. Briscese, G. Calcagni, L. Modesto and G. Nardelli,
\textit{Form factors, spectral and K\"all\'en-Lehmann representation in nonlocal quantum gravity.}
JHEP \textbf{08}  (2024) 204.




\bibitem{Roberts} C.D.~Roberts,
\textit{Hadron properties and Dyson-Schwinger equations,}
Prog. Part. Nucl. Phys. \textbf{61} (2008)  50, 
arXiv:0712.0633.

\bibitem{Widder} D.V. Widder,
\textit{Necessary and sufficient conditions for the representation of
a function by a doubly infinite Laplace integral,}
Bull. Amer. Math. Soc. 40 (1934) 321.

\bibitem{Krasnikov} N.V.~Krasnikov,
{\it Nonlocal gauge theories},
Theor. Math. Phys.  {\bf 73}  (1987) 1184.

\bibitem{Efimov} G.V.~Efimov,
\textit{Non-local quantum theory of the scalar field,}
Commun. Math. Phys. {\bf 5} (1967) 42. 

\bibitem{Ezquerro} M. Asorey, F. Ezquerro and I. L. Shapiro
\textit{Unitarity bounds in higher derivative field theories}
(In progress).

\bibitem{gclr} G. Calcagni and L. Rachwal,
\textit{Ultraviolet-complete quantum field
theories with fractional operators}
JCAP {\bf 09} (2023) 003


\bibitem{Simon-90} J.Z. Simon,
{\it Higher-derivative Lagrangians, nonlocality,
problems, and solutions,}
Phys. Rev. {\bf D41} (1990) 3720.

\bibitem{ParSim} L. Parker and J.Z. Simon,
{\it  Einstein equation with quantum corrections reduced
to second order,}
Phys. Rev. {\bf D47} (1993) 1339, 
gr-qc/9211002.

\bibitem{HD-Stab} F. de O. Salles and I.L. Shapiro,
{\it Do we have unitary and (super)renormalizable quantum
gravity below the Planck scale?}.
Phys. Rev. {\bf D89} 084054 (2014);
{\bf 90}, 129903 (2014) [Erratum],
\ arXiv:1401.4583.

\bibitem{PP} P.~Peter, F. de O. Salles and I.L.~Shapiro,
{\it On the ghost-induced instability on de Sitter background,}
Phys. Rev. {\bf D97} (2018) 064044,
arXiv:1801.00063.

\bibitem{Zimmermann:1987ne}
W.~Zimmermann,
MPI-PAE/PTh-61/87, https://inspirehep.net/literature/249522

\bibitem{Martin-2001}
J.~Martin and R.H.~Brandenberger,
\textit{The TransPlanckian problem of inflationary cosmology,}
Phys. Rev.  \textbf{D63} (2001) 123501,
hep-th/0005209;
J.~Martin and R.H.~Brandenberger,
\textit{On the dependence of the spectra of fluctuations in
inflationary cosmology on transPlanckian physics,}
Phys. Rev. \textbf{D68} (2003) 063513,
hep-th/0305161.

\bibitem{Starobinsky-2001}
A.A.~Starobinsky,
\textit{Robustness of the inflationary perturbation spectrum to
trans-Planckian physics,}
Pisma Zh. Eksp. Teor. Fiz. \textbf{73} (2001) 415, 
astro-ph/0104043.

\bibitem{Barbado-2011}
L.C.~Barbado, C.~Barcelo, L.J.~Garay and G.~Jannes,
\textit{The Trans-Planckian problem as a guiding principle,}
JHEP \textbf{11} (2011) 112,
arXiv:1109.3593.

\bibitem{Dvali} G. Dvali, S. Folkerts and C. Germani,
\ {\it Physics of trans-Planckian gravity,}
Phys. Rev. {\bf D84} (2011) 024039,  arXiv:1006.0984;
%
G. Dvali and C. Gomez,
{\it Black holes quantum $N$-portrait},
Fortschr. Phys. {\bf 61} (2013) 742, arXiv:1112.3359.

\bibitem{Sorella1} L.~Baulieu,  D.~Dudal, M.S.~Guimaraes,
M.Q.~Huber, S.P.~Sorella, N.~Vandersickel, and D.~Zwanziger,
\textit{Gribov horizon and $i$-particles: about a toy model and the
construction of physical operators,}
Phys. Rev. D \textbf{82} (2010) 025021, arXiv:0912.5153.

\bibitem{Sorella2} M.A.L.~Capri, D.~Dudal, M.S.~Guimaraes,
L.F.~Palhares, and S.P.~Sorella,
\textit{Physical spectrum from confined excitations in a
Yang-Mills-inspired toy model,}
Int. J. Mod. Phys. A \textbf{28} (2013) 1350034, arXiv:1208.5676.

\bibitem{Sorella3} S.P.~Sorella,
\textit{Gluon confinement, $i$-particles and BRST soft breaking,}
J. Phys. A \textbf{44} (2011) 135403, arXiv:1006.4500.

\end{thebibliography}
\end{document}